\documentstyle[12pt,psfig]{article}
\def\simgt{\rlap{\lower 3.5 pt\hbox{$\mathchar \sim$}}\raise 1pt \hbox {$>$}}
\def\simlt{\rlap{\lower 3.5 pt\hbox{$\mathchar \sim$}}\raise 1pt \hbox {$<$}}
\begin{document} 
%
%
\vspace{-20mm} 
\title{
{\LARGE Lattice QCD Calculation of Flavor Singlet Matrix
Elements and $N$-$N$ Scattering Lengths}
         \vspace*{5mm}
\author{Yoshinobu Kuramashi 
\vspace*{3mm}  
\\
   {\it     National Laboratory for High Energy Physics (KEK)}\\
   {\it     Tsukuba, Ibaraki 305, Japan}\\ 
}}

\date{}
\maketitle

\begin{abstract}

We report on our recent study of flavor singlet matrix elements 
calculated  with the wall source method without gauge fixing.	
Results are presented for the mass difference between
$\eta^{\prime}$ and pseudoscalar octet mesons,   
the $\pi$-$N$ $\sigma$ term and the proton axial vector matrix
elements. An exploratory calculation of the $N$-$N$ scattering 
lengths is also discussed including a phenomenological 
study with one-boson exchange potentials.

\end{abstract}

\section{Introduction}

Numerical simulations of lattice QCD have been successfully applied to 
calculate  a wide of variety of physically interesting observables.
However, the class of observables treated to date are mostly restricted to 
the flavor non-singlet 
quantities and little is known for the flavor singlet ones. 
The reason is the technical difficulty of calculating amplitudes involving 
disconnected quark loops required in the latter case.
We have found, however, that the method of wall source without gauge fixing, 
previously developed to deal with a similar computational difficulty with 
full hadron four-point functions\cite{pipi}, allows to overcome the
problem.  Using this method we have recently calculated 
the $\eta^{\prime}$ meson mass\cite{eta1,eta2}, 
the $\pi$-$N$ $\sigma$ term\cite{sigma} and the flavor
singlet axial vector matrix elements of proton\cite{pspin}. 
The first aim of this article is to present a summary of our findings 
on these quantities.

Calculation of hadron scattering lengths is a first step going beyond a 
study of 
static properties of hadrons based on lattice QCD.  Of particular 
phenomenological interest is the nucleon-nucleon scattering lengths 
since, unlike the pion-pion and pion-nucleon cases, their values are not 
constrained by chiral symmetry.  The second aim of this article is 
to present our study of the subject, including a one-boson exchange 
model calculation on the quark mass dependence of the scattering 
lengths\cite{nn,scattfull}.
 
All of our calculations are carried out in quenched QCD at
$\beta=5.7$ using the Wilson quark action with the hopping
parameters $K=0.150-0.168$ on a $12^3\times 20$, 
$16^3\times 20$ or $20^3\times 20$ 
lattice.  Gauge configurations are generated by the
pseudo heat-bath method at 1000 sweep intervals. Errors in the physical 
quantities are estimated by a jackknife procedure.
We correct quark field normalization by $\sqrt{1-3K/4K_c}$ with 
$K_c=0.1694$ and employ the tadpole-improved one-loop 
expressions\cite{tadpole}
for the renormalization factors of the quark bilinear operators with 
$\alpha_{\overline{MS}}(1/a)=0.2207$ for the coupling constant.

 
\section{$\eta^{\prime}$ meson mass}  
 
The large value of the flavor singlet $\eta^{\prime}$ meson mass in 
comparison to the masses of octet pseudoscalar mesons constitutes the 
well-known U(1) problem\cite{Weinberg}. The current view on 
the resolution of this problem is based on the $1/N_c$
argument\cite{WV}:  the mass splitting $m_0^2=m_{\eta^{\prime}}^2-m_8^2$
is generated through iteration of virtual quark loops
in the $\eta^{\prime}$ propagator, each of which gives a
factor $m_0^2/(p^2+m_8^2)$, and the U(1) anomaly is supposed 
to give a large value to $m_0$. Our aim is to evaluate
$m_0$ directly in lattice QCD  and
examine the relation between $m_0$ and the topological charge. 

In quenched QCD the $\eta^{\prime}$ propagator only consists of 
the one-quark-loop amplitude having a single pole $1/(p^2+m_8^2)$ and 
the two-quark-loop amplitude with a double pole  
$m_0^2/(p^2+m_8^2)^2$.  This means that
the mass splitting $m_0$ can be extracted from
the ratio of the two amplitudes defined by
\begin{equation} 
R(t)=\frac{\langle \eta^\prime (t) \eta^\prime (0) 
\rangle_{\rm 2-loop}}
{\langle \eta^\prime (t) \eta^\prime (0) \rangle_{\rm 1-loop}}
\approx {m^2_0 \over 2m_8}t +{\rm const.}, 
\label{eq:etaratio}
\end{equation}
where each amplitude is projected onto the zero momentum state. 

In order to evaluate the two-loop amplitude, we calculate 
the quark propagator $G({\bf n},t)=\sum_{({\bf
n^{\prime\prime}},t^{\prime\prime})}
G({\bf n},t;{\bf n^{\prime\prime}},t^{\prime\prime})$
obtained {\it with unit source at every space-time site without
gauge fixing}.  The expression
$\sum_{\bf n}{\rm Tr}[G({\bf n},0)\gamma_5] 
\sum_{\bf n^{\prime}} {\rm Tr}~[G^\dagger({\bf
n^\prime},t)\gamma_5]$ equals the two-loop amplitude 
aside from extra contributions of gauge-variant non-local terms.
However, these must cancel out in the ensemble average.
This technique has the advantage that only a single quark matrix
inversion is required for each gauge configuration, which
enables us to achieve high statistics.

Our calculations of the $\eta^{\prime}$-octet mass splitting
are made with four values of the hopping parameter, $K=0.164$, 
$0.165$, $0.1665$ on an $L=12$ lattice and $K=0.168$ on 
an $L=16$ lattice. For each hopping parameter we use 300,
240, 240 and 43 gauge configurations, respectively, to evaluate
one- and two-loop amplitudes. In the last case three
exceptional configurations are excluded, for which we fail to 
find the solution of the quark matrix inversion. 

\begin{figure}[t]
\hspace{20mm}
\psfig{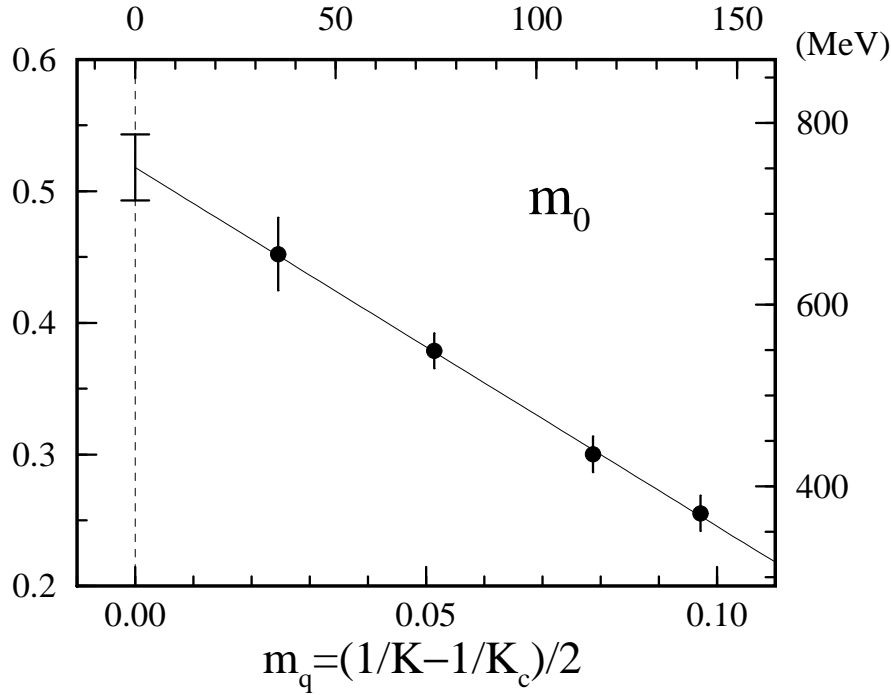} 
\vspace{-3mm}
\caption{\label{fig:m0atkc}
$m_0$ multiplied by $\protect\sqrt{N_f}=\protect\sqrt{3}$ as
a function  of $m_q=(1/K-1/K_c)/2$.  
Left ordinate and bottom abscissa are in
lattice units, while right ordinate 
and top abscissa are in physical units
using $a^{-1}=1.45$GeV.  Solid line is a linear fit 
which extrapolates to
$m_0=0.518(25)$ ( 751(39)MeV ) at $m_q=0$.}  
\end{figure}

We extract the mass splitting $m_0$ fitting the data of $R(t)$ 
to the linear form (\ref{eq:etaratio})
using the value of $m_8$ taken from a single exponential
fitting of the pion propagator (see Fig.~2 of Ref.~\cite{eta1}).
The results for $m_0$ multiplied by $\sqrt N_f=\sqrt 3$ to
obtain the physical value are plotted in Fig.~\ref{fig:m0atkc}
as a function of the quark mass $m_q=(1/K-1/K_c)/2$ using 
$K_c=0.1694$.  We observe a linear increase of $m_0$ 
as $m_q$ decreases toward the chiral limit. Employing a
linear extrapolation for $m_0$ to obtain the value at $m_q=0$,
we find $m_0=0.518(25)$ or $m_0=751(39)$MeV in physical
units with $a^{-1}=1.45(3)$GeV determined from the $\rho$
meson mass.

The value of $m_0$ we obtained may be compared with 
the empirical mass difference
$m_0^2=m_{\eta^{\prime}}^2-(4m_K^2+3m_{\pi}^2+m_{\eta}^2)/8\approx
(866{\rm MeV})^2$, which is the naive estimation ignoring a small 
mixing between $\eta$ and $\eta^{\prime}$. While a precise
agreement is not obtained, both values are sufficiently
close to conclude that the $\eta^{\prime}$-octet mass
splitting can be understood within quenched QCD.

Let us next consider the topological charge dependence 
of the $\eta^{\prime}$-octet mass splitting $m_0$.
The topological charge on the lattice is defined as
\begin{equation}
Q =\frac{1}{32\pi^2}  \sum_n \epsilon_{\mu\nu\rho\sigma}
{\rm ReTr}\{U_{\mu\nu}(n)U_{\rho\sigma}(n)\},
\label{eq:topologicalcharge}
\end{equation}
where $U_{\mu\nu}(n)$ is the plaquette in the $\mu\nu$ plane
at site $n$. We calculate $Q$ by the cooling 
procedure\cite{cooling} employing 25 cooling 
sweeps to wash out ultraviolet fluctuations.

To check the anomalous Ward identity
relation $m_0^2=2N_f\chi/f_\pi^2$ derived by Witten and Veneziano 
in the large $N_c$ limit\cite{WV},  we calculate
the topological susceptibility $\chi=\langle Q^2
\rangle /({\it space}$-${\it time\; volume})$
using our 300 pure gauge configurations on an $L=12$
lattice, and find $\chi=4.76(39)\times 10^{-4}$.
Combining with $f_\pi=0.0676(24)$ at $m_q=0$ obtained 
on the same set of gauge configurations used for the calculation of 
$m_0$, we find $\sqrt{2N_f\chi/f_\pi^2}=1146(67)$MeV for $N_f=3$ 
with the aid of $a^{-1}=1.45(3)$GeV. The result is somewhat higher 
compared to the value of $m_0$ estimated from the direct measurement.
It is possible that the disagreement arises from  
chiral symmetry breaking of the Wilson quark action which
induces extra terms in the anomalous Ward identity.

In order to further examine the relation between 
the $\eta^{\prime}$-octet mass splitting and
the topological charge we classify gauge configurations 
according to the value of $\vert \bar Q\vert$, which is the absolute magnitude 
of $Q$ rounded off to the nearest integer, and
evaluate $R(t;\vert \bar Q\vert )$ on each ensemble. 
For the one-loop amplitude in the denominator we use
the average over the whole ensemble since it shows little
variation depending on $\vert \bar Q\vert $.
The mass splitting $m_0$ is extracted by fitting the
data of $R(t;\vert \bar Q\vert )$ to the form 
of (\ref{eq:etaratio}).

\begin{figure}[t]
\hspace{17.5mm}
\psfig{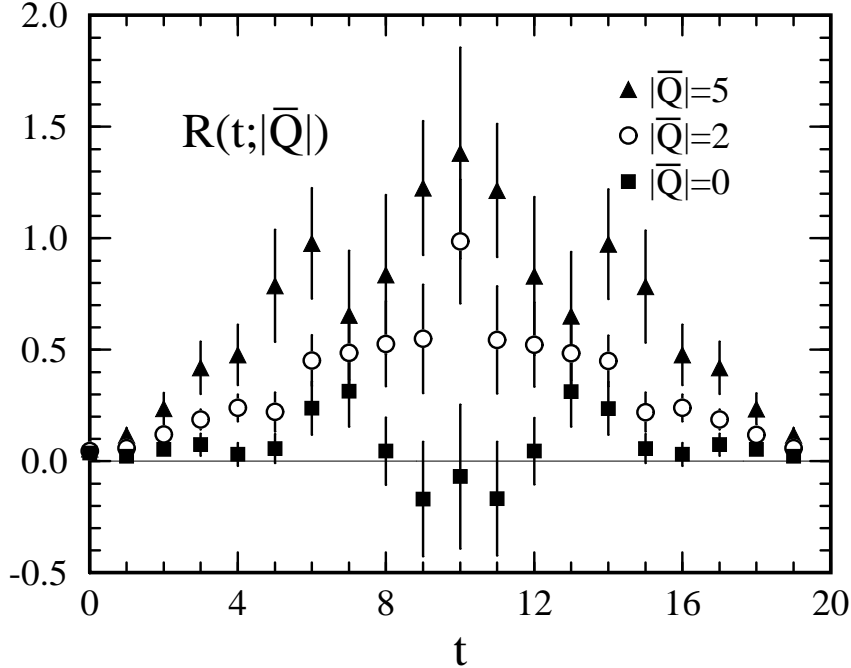} 
\vspace{-3mm}  
\caption{\label{fig:RtQ}
Representative values of the ratio $R(t;\vert{\bar Q}\vert)$ of two-
and single-quark loop contributions to 
the $\eta^{\prime}$ propagator for several values of 
$\vert {\bar Q}\vert $ at $K=0.1665$ on an $L=12$ lattice.}   
\end{figure}
 
In Fig.\ref{fig:RtQ} we present $R(t;\vert \bar Q\vert)$ at 
$K=0.1665$ for typical values of $\vert \bar Q\vert $: 
$\vert \bar Q\vert=0$, $2$, $5$. We see that the
function $R(t;\vert \bar Q\vert )$ is consistent with zero 
for $\vert \bar Q\vert =0$ and clearly increases as 
$\vert \bar Q\vert $ increases. This fact demonstrates 
that the excitation of instantons generates 
the contribution of the two-loop amplitude. 
The dependence of the $\eta^{\prime}$-octet mass 
splitting $m_0$ on $\vert \bar Q\vert $ is listed in
Table~\ref{tab:m0Q}. The increase of $m_0$
with $\vert \bar Q\vert $ shows that the $\eta^{\prime}$
meson mass increases with the winding numbers of
the instantons. 

In full QCD with dynamical quarks  
topological excitations become suppressed toward vanishing sea quark mass 
due to the fermionic zero modes.
It is an interesting problem to understand how the $\eta^\prime$ meson mass 
survives the suppression.  Full QCD simulations for examining this point 
have to overcome the difficulty that topological fluctuations are known to 
exhibit long-range correlations in Monte Carlo time\cite{fullTC}. 
               
\begin{table}[h]
\vspace*{-5mm}
\begin{center}
\caption{\label{tab:m0Q}
$\vert \bar Q \vert$ dependence of $m_0$ in lattice units 
at $K=0.1665$ on an $L=12$ lattice.}
\vspace*{2mm}
\begin{tabular}{c|cccccccc}
\hline
$\vert \stackrel{}{\bar Q} \vert$ 
& 0 & 1 & 2 & 3 & 4 & 5 & 6 & $\geq 7$  \\
$\sharp{\rm conf.}$ &25 & 47 & 51 & 20 & 35 & 22 & 12 & 28 \\
$m_0(\vert \stackrel{}{\bar Q} \vert)$ 
& 0.19(4) & 0.25(4) & 0.36(2) & 0.34(4)
& 0.34(3) & 0.49(4) & 0.45(6) & 0.57(4) \\
\hline
\end{tabular}   
\end{center} 
\vspace*{-5mm}
\end{table}
 

\section{Flavor singlet nucleon matrix elements}

\subsection{scalar matrix elements}
 
The $\pi$-$N$ $\sigma$ term is a fundamental quantity measuring 
the degree of chiral symmetry breaking in the
strong interaction and defined as $\sigma
=(m_{u}+m_{d})/2\cdot \langle N\vert{\bar u}u+{\bar
d}{d}\vert N\rangle$. Study of the $\sigma$ 
term has a long history and still shows some controversy 
due to the difficulty of estimating the sea quark contribution
including the strangeness content of the nucleon.  
A direct calculation of the $\sigma$ term
in lattice QCD potentially provides a thorough resolution
of this issue. Some previous calculations\cite{Gupta,sLiu} suggest
that the disconnected contribution is as large as the
connected one. However,
those calculations were made only at quite heavy quark masses 
and it seems difficult to extrapolate their result to the physical 
point (see Fig.~1 of Ref.~\cite{latt94}).

We have made a systematic calculation of the
quark mass dependence of the $\pi$-$N$ $\sigma$ term.
The nucleon matrix elements of scalar quark density are extracted from a 
linear fit in $t$ of the
ratio of nucleon three-point function divided by the
nucleon propagator\cite{threepoint}. 
Dirichlet boundary condition is 
employed in the temporal direction for quark propagators.
The source method\cite{source} is used for the
connected contribution. The disconnected quark loops are 
evaluated using quark propagators solved
with unit source for every space time site without gauge
fixing.

\begin{table}[t]
\vspace*{-5mm}
\begin{center}
\caption{\label{tab:scalar}
Nucleon scalar matrix elements corrected by  tadpole-improved 
$Z$ factor as a function of $K$ for up and down
quarks. $\langle N\vert \bar ss\vert N\rangle$ represents an
interpolated value at the physical strange quark mass 
$K_s=0.1648$ with up and down quark masses fixed.}
\vspace*{2mm}
\begin{tabular}{llllllll} 
\hline 
$K$  &
$L$& $\sharp$conf. &\multicolumn{2}{c}{$\stackrel{}{\langle N\vert \bar uu+\bar
dd\vert N\rangle}$} & $\langle N\vert \bar ss\vert N\rangle $ & $F_S$ & $D_S$ \\
     & &   & conn. & disc.  \\  
\hline
0.1600 & 12 & 300 & 2.323(15) & 3.56(76)  & 2.09(25) & 0.749(5)   
& $-$0.074(2)\\
       & 16 & 260 & 2.326(17) & 3.02(98)  & 1.78(32) & 0.749(7)  	
& $-$0.075(6)\\
0.1640 & 12 & 300 & 2.413(30) & 4.58(92)  & 2.36(29) & 0.762(10)  
& $-$0.126(6)\\ 
       & 16 & 260 & 2.378(30) & 4.1(1.2)  & 2.10(40) & 0.748(12) 	
& $-$0.128(11)\\
0.1665 & 12 & 400 & 2.693(81) & 5.1(1.1)  & 2.66(35) & 0.831(25)  
& $-$0.200(16)\\ 
       & 16 & 260 & 2.565(73) & 5.6(1.6)  & 2.70(51) & 0.783(26)	 
& $-$0.204(21)\\
\hline
$K_c$  & 12 & & 2.615(61) & 5.8(1.4)  & 2.84(44)  & 0.802(19)  &
$-$0.208(11)\\ 
       & 16 & & 2.515(60) & 6.2(1.9)  & 2.89(61)  & 0.765(23)  &
$-$0.222(20)\\
\hline
\end{tabular}
\end{center}
\vspace*{-5mm}
\end{table}               

In Table~\ref{tab:scalar} we summarize our results
for the nucleon scalar matrix elements for each spatial size 
$L$ and the hopping parameter $K$. Since the results
for $L=12$ and $16$ shows little finite size effects, 
we give the averages of $L=12$ and
$L=16$ results in the following.

For the isoscalar matrix element $\langle N\vert \bar uu+\bar
dd\vert N\rangle$ we observe that the disconnected contribution is about
twice larger than the connected one.  
In the the chiral limit we find for the ratio $\sigma_{\rm disc}/
\sigma_{\rm conn}=2.35(46)$.
 
Calculating the physical value of the $\sigma$ term requires an 
estimate of the quark mass.  For this purpose
hadron masses $m_\pi$, $m_\rho$, $m_N$ are calculated
using the same set of hopping parameters 
on the same set of gauge configurations. 
The light quark mass ${\hat m}=(m_u+m_d)/2=0.0034(1)$  
$(m_q=(1/K-1/K_c)/2)$
is determined using $m_\pi/m_\rho=0.18$
and the strange quark mass $m_s=0.0829(19)$ $(K_s=0.1648)$
from $m_K/m_\rho=0.64$. The lattice scale 
$a^{-1}=1.46(2)$GeV is fixed by $m_\rho$ at $K_c$.
Here we take the average of results for the $L=12$ and $16$ lattices
since we do not detect any finite size effects for hadron masses
between the two lattice sizes\cite{latt94}.   

We estimate the physical value of $\sigma$ in two ways.
The isoscalar matrix element at $K_c$ multiplied by 
the light quark mass ${\hat m}$ gives $\sigma=44(6)$MeV. 
Alternatively, the ratio $m_N\sigma/m_\pi^2$ extrapolated
linearly in $m_q$ to the chiral limit yields $60(9)$MeV with the
aid of experimental pion and nucleon masses.
The difference between the two values of $\sigma$ 
originates from the large ratio $m_N/m_\rho=1.47(3)$
in our simulation carried out at $\beta=5.7$. Within this uncertainty
these estimates show an encouraging agreement
with experimental estimates $\sigma=45$MeV\cite{GLS}.
For the strangeness content 
of nucleon we find a fairly large value for the $y$ parameter
$y=2\langle N\vert {\bar s}s\vert N\rangle 
/\langle N\vert {\bar u}u+{\bar d}d\vert N\rangle=0.66(13)$,
as compared to a phenomenological estimate $y\approx 0.2$\cite{GLS}.

Obvious systematic uncertainties in our results are scaling
violation and sea quark effects. In addition 
we should note the potential problem that the physical light quark
mass for the Wilson quark action substantially decreases toward
smaller lattice spacing in quenched QCD and that the full
QCD values are factor $2-3$ smaller than those for quenched
QCD\cite{Ukawa}.

Let us also make a remark concerning the flavor non-singlet $SU(3)$ reduced 
matrix elements $F_S$ and $D_S$.  These may be evaluated from the 
relations $F_S=\langle p\vert {\bar u}u-{\bar s}s\vert p\rangle/2$
and $D_S=\langle p\vert {\bar u}u-2{\bar d}d+{\bar s}s\vert p\rangle/2$
with $\vert p\rangle$ the proton state, where we may ignore disconnected 
contributions to leading order in flavor symmetry breaking.
For the mass splitting of the baryon octet, our results for the reduced 
matrix elements yield 
$m_\Xi-m_N=(m_s-\hat m)\cdot 2F_S=0.182(6)$GeV and 
$-3/2 (m_\Sigma-m_\Lambda)$=$(m_s-\hat m)\cdot 2D_S$=$-0.0498(30)$GeV.
The magnitude of these values is roughly a factor two
smaller compared to the experimental values $0.379$GeV and
$-0.116$GeV, while our result for the ratio $D_S/F_S=-0.275(16)$ is
reasonably consistent with the experimental estimate
$D_S/F_S\approx -0.32$.  A source of uncertainty, besides scaling violations 
and sea quark effects, is the validity of the first order approximation 
for the mass splitting.  
This point should be resolved by a simultaneous calculation  of the baryon 
mass splitting and scalar matrix elements, which we leave for future
analysis.
 

\subsection{axial vector matrix elements}
    
\begin{table}[b]
\setlength{\tabcolsep}{0.2pc}
\vspace*{-5mm}
\begin{center}
\caption{\label{tab:axial}
Proton axial vector matrix elements corrected by 
tadpole-improved $Z$ factor as a function of $K$ 
for up and down quarks. ${\Delta d}_{\rm disc}$ equals
${\Delta u}_{\rm disc}$. $\Delta s$ represents an
interpolated value at the physical strange quark mass 
$K_s=0.1648$ with up and down quark masses fixed.}
\vspace*{2mm}
\begin{tabular}{llllllll} 
\hline
$K$ & $L$ & ${\Delta u}_{\rm conn}$ & ${\Delta d}_{\rm conn}$ 
& ${\Delta u}_{\rm disc}$  
& ${\Delta s}$ & $F_{A}$ & $D_{A}$\\
\hline
$0.1600\;$  & $12\;$ & 0.8840(60) & $-0.2458(23)$ & 
                     &            & $0.4420(30)$  & $0.6878(46)$ \\   
            & $16\;$ & 0.9071(92) & $-0.2470(35)$ & $-0.025(10)$ 
                     & $-0.0374(89)$ & $0.4536(46)$ & $0.7011(37)$ \\   
$0.1640\;$  & $12\;$ & 0.795(11)  & $-0.2387(51)$ &          
                     &            & $0.3974(55)$  & $0.6358(83)$ \\   
            & $16\;$ & 0.839(19)  & $-0.2382(87)$ &$-0.066(23)$ 
                     & $-0.069(16)$ & $0.4196(93)$ & $0.6580(71)$ \\   
$0.1665\;$  & $12\;$ & 0.761(20)  & $-0.232(11)$  & 
                     &            & $0.3806(98)$  & $0.611(15)$ \\   
            & $16\;$ & 0.818(39)  & $-0.231(23)$  &$-0.093(54)$  
                     & $-0.087(33)$ & $0.409(20)$ & $0.641(16)$  \\   
\hline
$K_c$       & $12\;$ & 0.694(20)  & $-0.2284(96)$ & 
                     &            & $0.347(10)$   & $0.575(15)$ \\   
            & $16\;$ & 0.763(35)  & $-0.226(17)$  & $-0.119(44)$  
                     & $-0.109(30)$ & $0.382(18)$ & $0.607(14)$ \\   
\hline
\end{tabular}
\end{center} 
\vspace*{-5mm}
\end{table}
 
Interest in the flavor singlet axial vector matrix elements
of proton was inspired by the EMC result 
for the spin dependent structure function $g_1$ of proton\cite{EMC}, 
which indicated that the fraction of proton spin carried by
quarks has a small value and that the strange quark
contribution is unexpectedly large and negative.  
Recent analyses combining data on proton\cite{EMC,pexp,E143}
, deuteron\cite{E143,dexp} and neutron\cite{nexp} targets 
yield $\Delta \Sigma=\Delta u+\Delta
d+\Delta s=+0.83(3)-0.43(3)-0.10(3)=0.31(7)$ at the
renormalization point $\mu^2=10\mbox{GeV}^2$\cite{EK} or 
$+0.832(15)-0.425(15)-0.097(18)=0.31(4)$ at
$\mu^2=\infty$\cite{AR}, where 
$\Delta q$ is the forward matrix element of
$\bar q\gamma_i\gamma_5 q$ for a proton polarized in the
$i$-th direction averaged over $i=1$, $2$, $3$.
These results are far from the naive constituent quark
model which predict $\Delta u=4/3, \Delta d=-1/3$ and $\Delta s=0$. 

We have attempted a lattice calculation of $\Delta q$ for each flavor $q$ 
including both connected and disconnected
contributions. 
We evaluate the proton matrix elements of the axial vector
current employing the same calculational technique,
the same set of gauge configurations and the same hopping
parameters as for the case of the nucleon scalar matrix elements.    
               
\begin{figure}[t]
\hspace{20mm}
\psfig{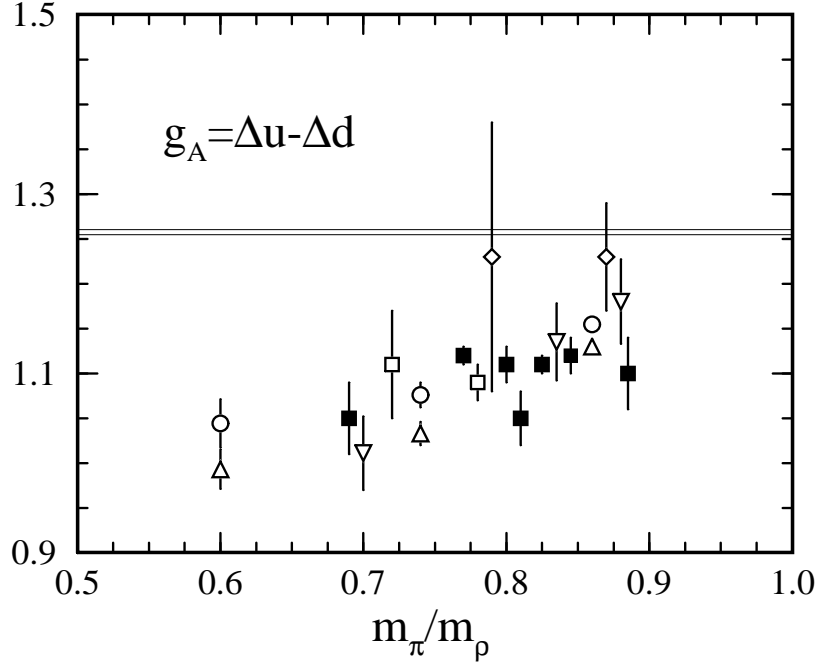}
\vspace{-3mm}
\caption{\label{fig:fullga}
Axial coupling $g_A$ for quenched (open) and full QCD
(closed) with Wilson quark action. Our results for $L=16$ 
(circles) and $L=12$ (up-triangles) are at $\beta=5.7$. 
$\beta=6.0$ for other quenched results
(open squares\protect\cite{Gupta},
down-triangles\protect\cite{DESY}, 
diamonds\protect\cite{gaLiu},).
Full QCD results (closed squares\protect\cite{Gupta}) 
are at $\beta=5.4-5.6$.
Horizontal lines represent the upper and lower
bound of $1\sigma$ for the experimental value
$g_A=1.2573\pm 0.0028$.}
\end{figure}     
                            
Our results are summarized in Table~\ref{tab:axial}.
We observe little finite size effects for the connected
contributions. 
For the disconnected piece statistical
fluctuations turned out to be significantly larger than 
those for the scalar matrix elements, to the extent that
a reliable signal was not obtained on an $L=12$ lattice.
  
Our results for $L=16$ show that the disconnected
contribution, albeit involving a substantial error of
$50\%$, are negative and that their magnitude
is small compared to those for the connected ones.  
Also, the disconnected contributions slightly
increase toward the chiral limit while the connected
ones exhibit a slight decrease.   
Adding both contributions for each flavor and 
making a linear extrapolation to $K=K_c$, we find   
$\Delta \Sigma =\Delta u+\Delta d+\Delta s=+0.638(54)
-0.347(46)-0.109(30)=+0.18(10)$ with the disconnected 
contribution to $\Delta u$ and $\Delta d$ equal to
$-0.119(44)$.
These values show a reasonable agreement with the 
phenomenological estimates quoted above\cite{EK,AR}. 

In Table~\ref{tab:axial} we also list the flavor non-singlet
matrix elements $F_A=(\Delta u-\Delta 
s)/2$ and $D_A=(\Delta u-2\Delta d+\Delta s)/2$ 
evaluated without the disconnected contributions under an
assumption of exact SU(3) flavor symmetry. 
In the chiral limit we find $F_A=0.382(18)$
and $D_A=0.607(14)$ for $L=16$, which give the axial charge 
$g_A=F_A+D_A=0.985(25)$
and the ratio $F_A/D_A=0.629(33)$. Compared to the experimental
values $g_A=1.2573(28)$ and $F_A/D_A=0.586(19)$\cite{PDG}, 
the ratio shows a good agreement, while $g_A$ is about $25\%$ 
smaller. 
Other calculations of $g_A$ for quenched QCD
at $\beta=6.0$\cite{Gupta,DESY,gaLiu} and for full
QCD\cite{Gupta} at $\beta=5.4-5.6$,
with the lattice spacing of $a\approx 0.15-0.1$fm, yield
similar results if we use the same renormalization factor
as we employed. We illustrate this point in Fig.~\ref{fig:fullga}.
The agreement of results for a variety of simulation parameters indicates 
that possible systematic errors such as finite size
effects, scaling violations and quenching effects 
are too small to explain the deviation of the lattice estimates 
from the experimental value. The origin of the discrepancy  
is an open problem which should be resolved in future work.
  

\section{$N$-$N$ scattering lengths}
               
\begin{figure}[p]
\hspace{20mm}
\psfig{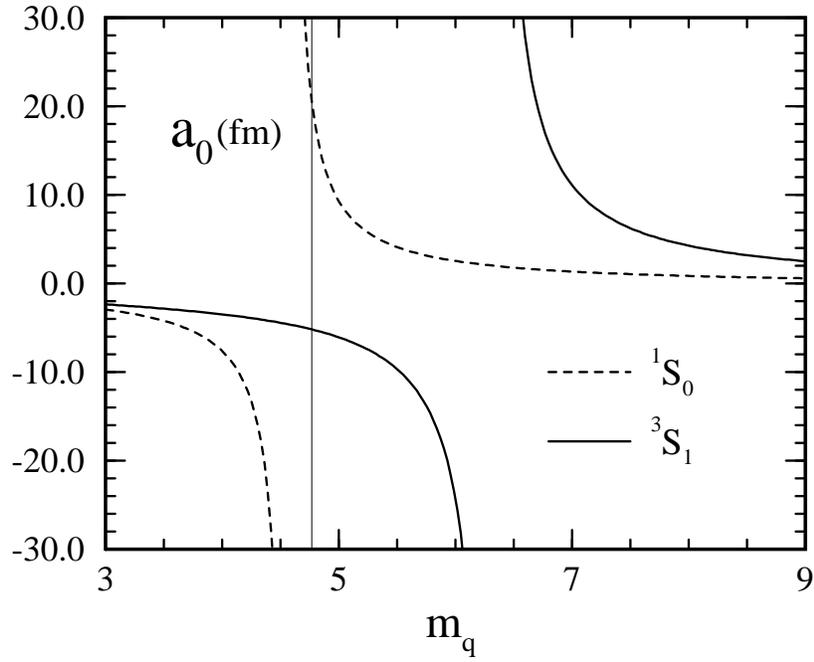}
\vspace{-3mm}
\caption{\label{fig:nnobe} 
Quark mass $m_q$ dependence of $N$-$N$ scattering lengths based
on a model of one-boson exchange potentials. Vertical
line represents the physical quark mass $m_q=4.8$MeV.}
\end{figure}                   
               
\begin{figure}[p]
\hspace{20mm}
\psfig{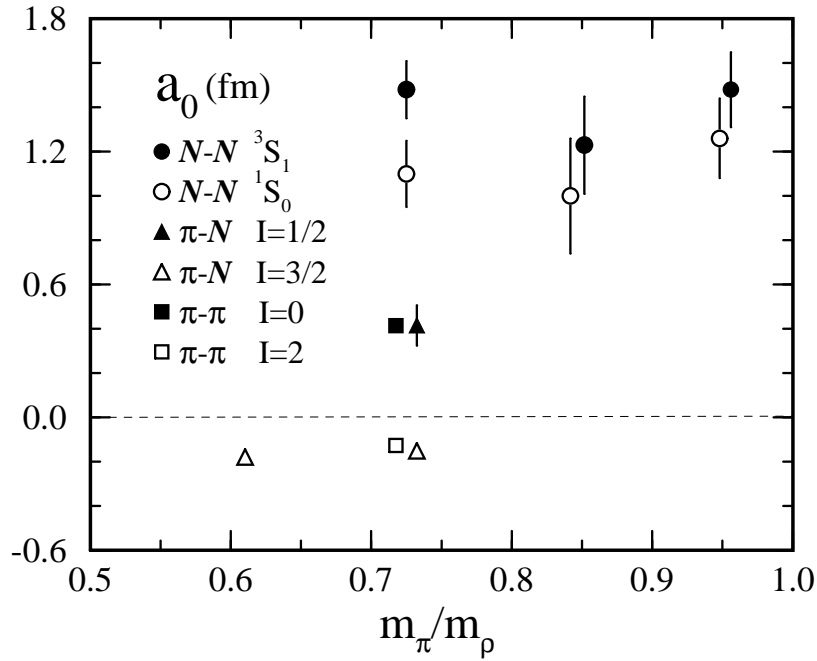}
\vspace{-3mm}
\caption{\label{fig:scatall} 
$N$-$N$ scattering lengths in units of fm as compared to $\pi$-$N$ and
$\pi$-$\pi$ scattering lengths\protect\cite{scattfull} in quenched QCD 
at $\beta=5.7$ calculated with
the Wilson quark action.  Conversion to physical units 
is made with $a=0.137(2)$fm
determined from the $\rho$ meson mass.}
\end{figure}                   

Lattice calculation of hadron scattering 
lengths is based on the
L\"{u}scher's formula\cite{Luescher} which relates the $s$-wave 
scattering lengths $a_0$ to the energy shift of two 
particle state at zero relative momentum confined 
in a finite periodic spatial box of a size $L^3$:     
\begin{equation} 
E-(m_1+m_2)=-\frac{2\pi
(m_1+m_2) a_0}{m_1 m_2 L^3}\left(1+c_1\left(\frac{a_0}{L}\right)
+c_{2}\left(\frac{a_0}{L}\right)^2\right)+O(L^{-6}) 
\label{eq:Luescher} 
\end{equation}  
with $c_1=-2.837297$, $c_2=6.375183$. 
For the nucleon-nucleon case application of this formula faces the difficulty 
that the experimental values $a_0({}^3S_1)=-5.432(5)$fm
and $a_0({}^1S_0)=+20.1(4)$fm\cite{nnexp} are quite large so that a large 
lattice is required to suppress $O(L^{-6})$ corrections in (\ref{eq:Luescher}). 
The negative sign in the
${}^3S_1$ channel due to the presence 
of the deuteron bound state (Levinson's theorem) brings an additional
complication that the energy of the lowest scattering state
orthogonal to the bound-state deuteron has to be computed to
apply (\ref{eq:Luescher}).
 
To avoid this complication we take a strategy to start our
simulation from the heavy quark mass region.  In order to have an idea 
on the behavior of $N$-$N$ interactions toward large quark 
masses, we carry out a phenomenological analysis employing a model of 
one-boson exchange potentials. Since these models include a large
number of parameters fine-tuned to reproduce the
experimental results for the scattering phase shifts,
we make this study not to analyze detailed quark mass dependence
of scattering lengths, but to obtain a rough feature of the
nuclear force under variation of quark mass around
the physical one.  

Specifically we employ the model of Ref.~\cite{OBEP}
and vary the value of $m_{\pi}$ according
to our lattice results for the slope $m_{\pi}^2/m_q$.
The mass of the scalar particle $\sigma$ is also varied 
in proportion to $m_{\pi}$ since
$\sigma$ represents the contribution of two-pion 
exchange. For $N$, $\rho$ and $\omega$ masses we assume the linear
$m_q$ dependence $m_h=a_h+b_h\cdot m_q$, fixing the slope 
$b_h$ with the aid of our lattice results.
Other meson masses $m_\eta$, $m_\delta$, $m_\phi$
and the meson-nucleon coupling constants are fixed to the 
the physical value. 

The quark mass dependence of the $N$-$N$ scattering lengths  
obtained in this way is shown in Fig.~\ref{fig:nnobe}.
We find a divergence of the scattering length in the ${}^3S_1$
channel taking place at $m_q=6.3$MeV that signals unbinding
of the deuteron, which is only 30\% larger than the physical point 
$m_q=4.8$MeV. On the other hand the ${}^1S_0$ channel forms the
bound state at a slightly smaller quark mass of $m_q=4.6$MeV.
For both channels the magnitude 
of the scattering length reduces quickly for heavier quark and becomes 
close to the order of 1~fm.
It is reasonable to expect that the scattering lengths
take values similar to the hadron size in the absence
of bound-state effects.

These results suggest that the deuteron becomes unbound as the quark mass 
increases, and hence L\"uscher's formula can be applied to extract the 
$N$-$N$ scattering lengths for heavy quark.  
Our lattice study of the $N$-$N$ scattering lengths is
carried out based on this consideration using the same technique as we 
employed for the $\pi$-$\pi$
and $\pi$-$N$ cases\cite{nn,scattfull}. 
For the calculation of the nucleon four-point function 
gauge configurations are fixed to the Coulomb gauge over all
space-time to enhance signals. We analyzed 20, 30 and 20
gauge configurations for the hopping parameters $K=0.150$,
$0.160$ and $0.164$, respectively, on a $20^3\times 20$ lattice.
In Fig.~\ref{fig:scatall} we compare the
results for the $N$-$N$ scattering lengths with those for 
$\pi$-$\pi$ and $\pi$-$N$ cases. 
As expected, the $N$-$N$ scattering 
lengths are of the order of 1~fm with a positive sign 
for both eigenchannels. They are substantially larger 
than the $\pi$-$\pi$ and $\pi$-$N$
scattering lengths already for a heavy quark corresponding
to $m_\pi/m_\rho\approx 0.74$. Also we note the
trend, albeit with sizable errors, that the values
for the ${}^3S_1$ channel are larger than those for the 
${}^1S_0$ channel.
This indicates a stronger attraction 
in the ${}^3S_1$ channel, consistent with the formation  
of the deuteron bound state at the physical quark mass where a bound state
is not yet formed in the  ${}^1S_0$ channel.    
 
If we reduce the quark mass toward the physical
point, we expect an increase of the $N$-$N$ scattering lengths
in both eigenchannels. For the ${}^3S_1$ channel the
scattering length should diverge at some quark mass signalling 
that the deuteron bound state is formed. However, 
actually observing such a phenomena will be difficult.  Large lattice sizes 
far exceeding the size of $2-3$fm, which
are accessible in current lattice QCD simulations, will be needed.  
Furthermore increase of statistical fluctuations for large times 
in the $N$-$N$ four-point functions is expected to become quite 
severe toward the chiral limit as shown
in Fig.~23 of Ref.~\cite{scattfull}. 
   

\section{Conclusions}

The method of wall source without gauge fixing has proven to be an 
effective tool for exploring the role of disconnected quark loop 
amplitudes in the flavor singlet sector of strong interactions.  
Our results for the $\eta^{\prime}$-octet mass splitting demonstrate 
that topologically non-trivial gauge configurations boost the $\eta^\prime$ 
meson mass to the level observed experimentally.  
Our application of the method to flavor singlet nucleon matrix elements 
have shown that disconnected contributions are substantial for the 
$\pi$-$N$ sigma term and the quark content of proton spin.  
The total contribution we found for these quantities are  
reasonably consistent with the experimental estimates.  
On the other hand, the flavor non-singlet pieces 
calculated from the connected contributions show 
substantial discrepancies between the lattice estimates 
and experiments. 
 
We have also described our exploratory calculation 
of the $N$-$N$ scattering lengths at heavy quark masses, for which 
we found results expected from considerations based on a 
phenomenological model of one-boson exchange potentials.

\end{document}